\documentclass[final]{svjour2}
\usepackage{graphicx}
\usepackage{rotating}
\usepackage{amssymb}
\usepackage{mathptmx}
\usepackage[numbers]{natbib}
\usepackage{subfigure}
\usepackage{wrapfig}
\makeatletter
\journalname{Journal of Low Temperature Physics}

\bibpunct{}{}{,}{s}{}{,}

\begin{document}

\newcommand{\hdblarrow}{H\makebox[0.9ex][l]{$\downdownarrows$}-}
\title{Development of readout electronics for \textsc{POLARBEAR-2} Cosmic Microwave Background experiment}

\author{K. Hattori$^{a}$ \and Y. Akiba$^{b}$ \and K. Arnold$^{c}$ 
\and D. Barron$^{d}$ \and A. N. Bender$^{e}$ \and A. Cukierman$^{d}$ \and T. de Haan$^{d}$ \and M. Dobbs$^{f}$
\and T. Elleflot$^{g}$ \and M. Hasegawa$^{b}$ \and M. Hazumi$^{b}$
\and W. Holzapfel$^{d}$ \and Y. Hori$^{d}$ 
\and B. Keating$^{g}$ \and A. Kusaka$^{h}$
\and A. Lee$^{d}$ \and J. Montgomery$^{f}$ \and K. Rotermund$^{i}$
\and I. Shirley$^{d}$ \and A. Suzuki$^{d}$ \and N. Whitehorn$^{d}$
}

\institute{
$^{a}$ Kavli IPMU (WPI), UTIAS, The University of Tokyo, Kashiwa, Chiba 277-8583, Japan\\
\email{khattori@berkeley.edu}\\
$^{b}$ High Energy Accelerator Research Organization (KEK), Tsukuba, Ibaraki 305-0801, Japan\\
$^{c}$ Department of Physics, University of Wisconsin, Madison WI 53706, USA\\
$^{d}$ Department of Physics, University of California, Berkeley, CA 94720, USA\\
$^{e}$ High Energy Physics Division, Argonne National Laboratory, Argonne, IL, USA 60439\\
$^{f}$ Physics Department, McGill University, Montreal, QC H3A 0G4, Canada\\
$^{g}$ Department of Physics, University of California, San Diego, CA 92093-0424, USA\\
$^{h}$ Physics Division, Lawrence Berkeley National Laboratory, Berkeley, CA 94720, USA\\
$^{i}$ Department of Physics and Atmospheric Science, Dalhousie University, Halifax, NS, B3H 4R2, Canada
}

\maketitle

\begin{abstract}

The readout of transition-edge sensor (TES) bolometers with a large multiplexing factor is
key for the next generation Cosmic Microwave Background (CMB) experiment, \textsc{Polarbear}-2 \cite{ref:PB2_overview},
having 7,588 TES bolometers.
To enable the large arrays, we have been developing a readout system with a multiplexing factor of 40
in the frequency domain.
Extending that architecture to 40 bolometers requires an increase in the bandwidth of the SQUID electronics,
above 4 MHz.
This paper focuses on cryogenic readout and shows how it affects cross talk and the responsivity of the TES bolometers.
A series resistance, such as equivalent series resistance (ESR) of capacitors for LC filters,
leads to non-linear response of the bolometers.
A wiring inductance modulates a voltage across the bolometers and causes cross talk.
They should be controlled well to reduce systematic errors in CMB observations.
We have been developing a cryogenic readout with a low series impedance and have tuned bolometers in the middle of
their transition at a high frequency ($>$ 3 MHz).


\end{abstract}

\section{Introduction}
For the next generation Cosmic Microwave Background (CMB) experiment, \textsc{Polarbear}-2 \cite{ref:PB2_overview}, 
we will build a receiver that has 7,588 TES bolometers coupled to two-band (95 and 150 GHz) polarization-sensitive antennas.
The kilopixel arrays are necessary to achieve the required sensitivity and our science goals. 
\textsc{Polarbear}-2 is an upgraded experiment based on \textsc{Polarbear}-1, 
which has measured the B-mode power spectrum \cite{ref:PB1-3}, the deflection power spectrum \cite{ref:PB1-1}
and its cross-correlation with cosmic infrared background \cite{ref:PB1-2}. 
\textsc{Polarbear}-1 is using a readout system with a multiplexing factor of 8 in the frequency domain, 
with the bolometers biased up to 1 MHz.
\textsc{Polarbear}-1 is using analog flux-locked loop (FLL) to read out the current through the SQUID amplifier.
 
\begin{figure}
\begin{center}
\includegraphics[%
  width=0.85\linewidth, bb=0 0 794 396
  ]{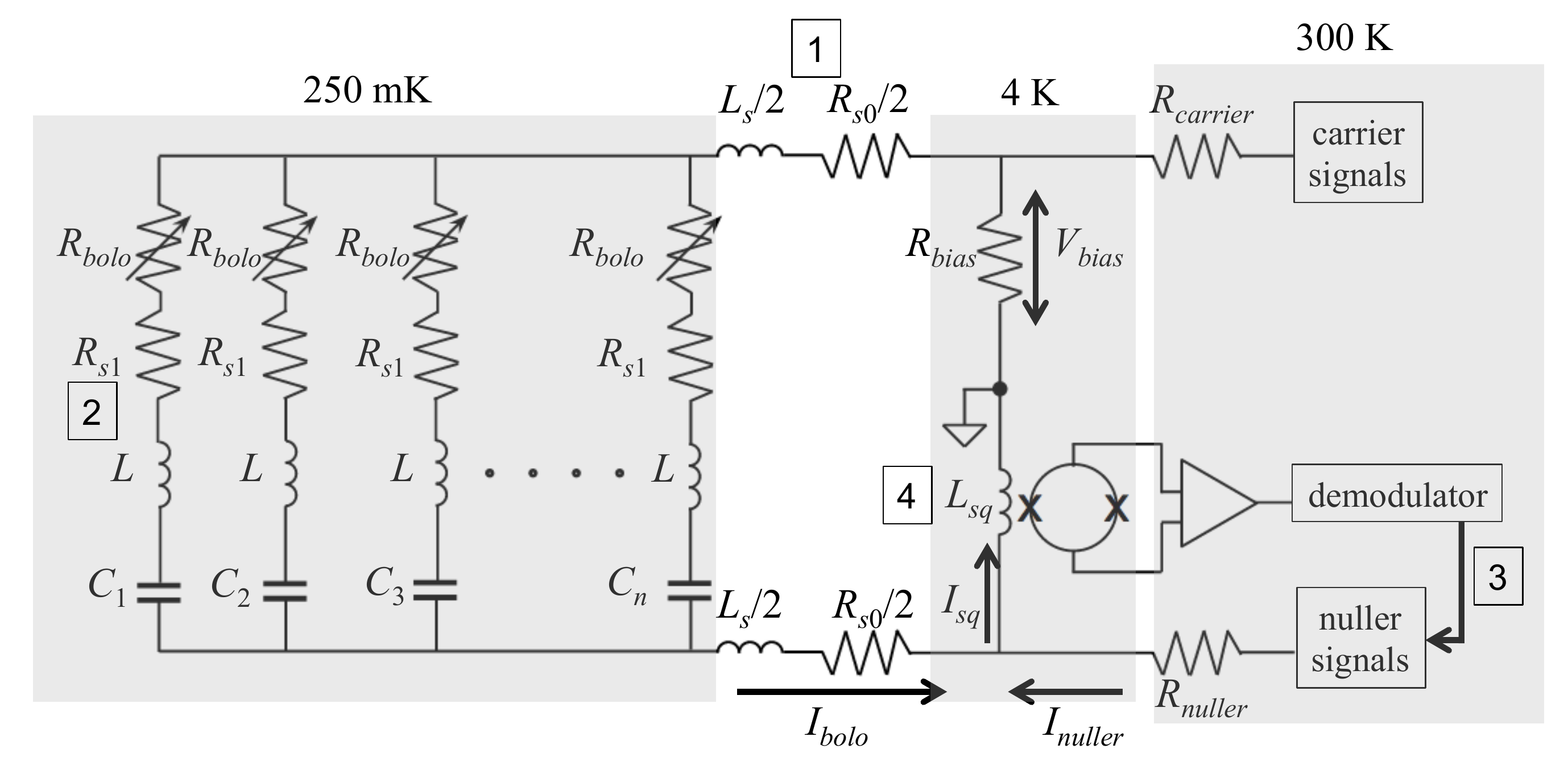}
\end{center}
\caption{Block diagrams showing the digital frequency-domain multiplexing readout of \textsc{Polarbear}-2.
(1) A cold wire connecting the bias resistor (4 K) and the LCR circuit (250 mK)
has a series inductance $L_s$ and resistance $R_{s0}$.
(2) Each LCR leg has an Equivalent Series Resistance (ESR) of a capacitor for an LC filter, $R_{s1}$. 
(3) Digital Active Nulling (DAN) on the digital frequency multiplexing platform is used
to increase the readout bandwidth.
(4) Current flowing through the SQUID input coil is nulled by DAN, such that $I_{sq} = I_{bolo} + I_{null} = 0$.
}
\label{fig:circuit_diagram}
\end{figure}

The \textsc{Polarbear}-2 focal plane is an order of magnitude larger than currently deployed TES arrays.
This requires a large multiplexing factor due to limitations on thermal loading from signal cables.
We are developing readout electronics, which multiplexes 40 TES bolometers through a single superconducting quantum interface device (SQUID). 
Extending that architecture to 40 bolometers requires an increase in the bandwidth of the SQUID electronics,
above 4 MHz. 
Since the analog FLL limits the usable bandwidth up to 1 MHz,
we use Digital Active Nulling (DAN) \cite{ref:DAN} \cite{ref:Amy_SPIE} 
on the digital frequency-domain multiplexing platform for multiplexing 40 bolometers,
as shown in Fig. \ref{fig:circuit_diagram}. 
With DAN, digital feedback is calculated for each bolometer, extending the useful bandwidth of the SQUID amplifier.
In this paper, we will describe cryogenic readout electronics, LC filters and cold cables,
which are also important for achieving the high multiplexing factor.
We are developing 40-channel LC filters and low-inductance superconducting cables. 
Detailed descriptions of the LC filters are found in \cite{ref:Kaja_LTD}.
This paper gives an overview of the cryogenic readout and its requirements for CMB observations.
Using the new readout components and the modified wiring, we will show the performance of bolometers biased at high frequency ($>$ 3 MHz).

\section{Cryogenic readout electronics}
Since the resistance of a TES bolometer is small, 
it is important to design the cryogenic readout electronics carefully,
and suppress the series impedance to voltage-bias the TES bolometers.  
In the following sections, we will give an overview of requirements placed on the series impedance.

\subsection{Series resistance}
A component that mainly contributes to the series resistance is an equivalent series resistance (ESR) of a capacitor
forming an LC filter.
As dielectric loss in the capacitors become larger at a higher frequency,
controlling the ESR becomes more important for a readout with a large multiplexing factor,
which requires a wide bandwidth.
The series resistance causes instability in electrothermal feedback
when $R_s > R_{bolo}$ with loop gain $>> 1$,
where $R_s$ is a series resistance, and $R_{bolo}$ is a TES resistance \cite{ref:Kent_Irwin}.
Even with $R_s < R_{bolo}$, a non-linear response of the TES bolometer is caused
and can create systematic errors with CMB observations.
Assuming that $\beta = \frac{\partial \log R}{\partial \log I} << 1$ for our TES bolometers, the non-linear response 
of the voltage-biased bolometer at a frequency, $\omega << R_{bolo}/L$,

\begin{equation}
\frac{dI}{dP} = -\frac{1}{I(R_{bolo} - R_s)} 
\frac{{\cal {L}}}{{\cal{L}} + \frac{R_{bolo} + R_s}{R_{bolo} - R_s}}
\frac{1}{1 + \frac{i\omega \tau}{\frac{R_{bolo} - R_s}{R_{bolo} + R_s}{\cal{L}} + 1}},
\label{eq:responsivity}
\end{equation}

where $I$ is the current through the TES, $P$ is input power, 
$\cal {L}$ is the loop gain, and $\tau$ is the time constant at ${\cal {L}} = 0$. 
This equation can be derived using equations shown in \cite{ref:Kent_Irwin}.
The equation is valid for CMB observations since we only use frequencies where $\omega << R_{bolo}/L$ for data analysis.
If the loop gain ${\cal {L}}$ is infinite and the series resistance
$R_s$ is zero, the gain $dI/dP$ is constant and the bolometer response is linear.
The series resistance varies the gain as the input power $P$ changes.
Considering the first order of Taylor series, the change in current as a function of input power is


\begin{equation}
\Delta I = \frac{dI}{dP}\Bigr| _{\Delta P = 0} \Bigl[
1 - \Bigl[
\frac{2R_s}{I^2(R_{bolo} - R_s)^2}
- \frac{1}{{\cal{L}}^2} \frac{d{\cal{L}}}{dP} \frac{R_{bolo} + R_s}{R_{bolo} - R_s} \frac{{\cal{L}}}{{\cal{L}} + \frac{R_{bolo} + R_s}{R_{bolo} - R_s}}
+ \frac{1}{{\cal{L}}} \frac{2R_s(R_{bolo} + R_s)}{(R_{bolo} - R_s)^3 I^2} \frac{{\cal{L}}}{{\cal{L}} + \frac{R_{bolo} + R_s}{R_{bolo} - R_s}}
\Bigr]\Bigr| _{\Delta P = 0}
\Delta P \Bigr],
\label{eq:taylor_series2}
\end{equation}

We used a constant-voltage biased condition, $dR_{bolo} / dI = -(R_{bolo} + R_s) / I$,
to derive Eq. \ref{eq:taylor_series2}.
The first term expresses linear response and other terms express the non-linear response.
When the loop gain is infinite and the series resistance is zero, the non-linear terms
become zero and the response is linear. 
The second and the last terms in Eq. \ref{eq:taylor_series2} are proportional to the series resistance.
The second term cannot be mitigated even with the infinite loop gain.
Our goal is $R_s / R_{bolo} \lesssim 0.2$, achieved by the \textsc{Polarbear}-1 readout,
to maintain the non-linearity in the same level.

Components that contribute to the series resistance are
(1) Equivalent Series Resistance (ESR) of a capacitor for an LC filter \cite{ref:Kaja_LTD}
(2) flex cables with copper lines which connect the wafer and the LC filters
(3) contact resistance of connectors (4) copper traces on printed circuit boards where SQUIDs are mounted
(5) a resistor biasing the bolometers (30 m$\Omega$).
In Fig. \ref{fig:circuit_diagram}, the resistive bridge consisting of $R_{carrier}$ and $R_{bias}$,
where $R_{carrier} >> R_{bias}$,
is equivalent to a voltage-source with $R_{bias}$ in series according to Thevenin's theorem.
Since readout lines on the wafer and cables between the LC boards and the SQUID mounting boards 
are superconducting, they don't contribute to the series resistance.
In Sec. \ref{sec:testing}, we present measurements of the total series resistance.

\subsection{Series inductance}
Though the series inductance can also break the constant voltage-biasing condition, 
most of the inductance in series with the bolometer can be mitigated and doesn't
appear in Eq. \ref{eq:responsivity}.
Most of the series inductance can be cancelled by
choosing the resonant frequency of the LC filter shifted by the series inductance.
However, contribution to cross talk between neighboring channels in frequency space \cite{ref:Matt_afMUX}
cannot be mitigated.
The change in the current through the series inductance $L_s$ in Fig. \ref{fig:circuit_diagram}
modulates the voltage across the entire LCR circuit, 
deposits electronic power to the bolometers and leads to cross talk.
The series resistance $R_{s0}$ shown in Fig. \ref{fig:circuit_diagram} also contributes to
the voltage modulation. At high frequency, the series impedance $i\omega L_s$ is
much higher than $R_{s0}$, and the series resistance is negligible.
Note that the current through the SQUID input coil is cancelled by DAN, 
such that $I_{sq} = I_{bolo} + I_{null} = 0$,
and the SQUID input impedance $i\omega L_{sq}$ doesn't contribute to cross talk.

The electrical cross talk should be smaller than the $\simeq$ 1\% 
specification set by the size of the CMB signals being measured. 
We chose logarithmic spacing to maintain cross talk across the entire readout bandwidth \cite{ref:hattori_SPIE}.
We set targeted series inductance to 60 nH, considering the length of the cold cables (about 60 cm)
and an achievable inductance (about 1 nH / cm).
Based on the series inductance and the acceptable cross talk, the peak frequencies of
the LC filters are from 1.6 MHz to 4.2 MHz. The frequency schedule allows a margin of error of resonant frequencies
of the LC filters.

\section{Cold readout with TES bolometers}
\label{sec:testing}

In this section, we present the evaluations of our cryogenic readout in a laboratory fMUX setup, shown in Fig. \ref{fig:circuit_diagram}.
We used a low inductance cable to connect the LCR circuit (0.3 K) to the bias resistor (4 K).
The cable consists of two segments: a low-inductance broadside-coupled stripline 
which has copper lines and high thermal conductivity, and a NbTi stripline with a similar geometry. 
The NbTi stripline was used to isolate the 4 K and 0.3 K stages thermally.
The total inductance for the cryogenic cable was $<$ 90 nH.
We are developing a NbTi stripline with a thinner substrate to achieve the targeted series inductance, 60 nH.

\begin{figure}
\begin{center}
\subfigure[At 0.6 K where the TESes are normal.]{
\includegraphics[height=4.2cm]{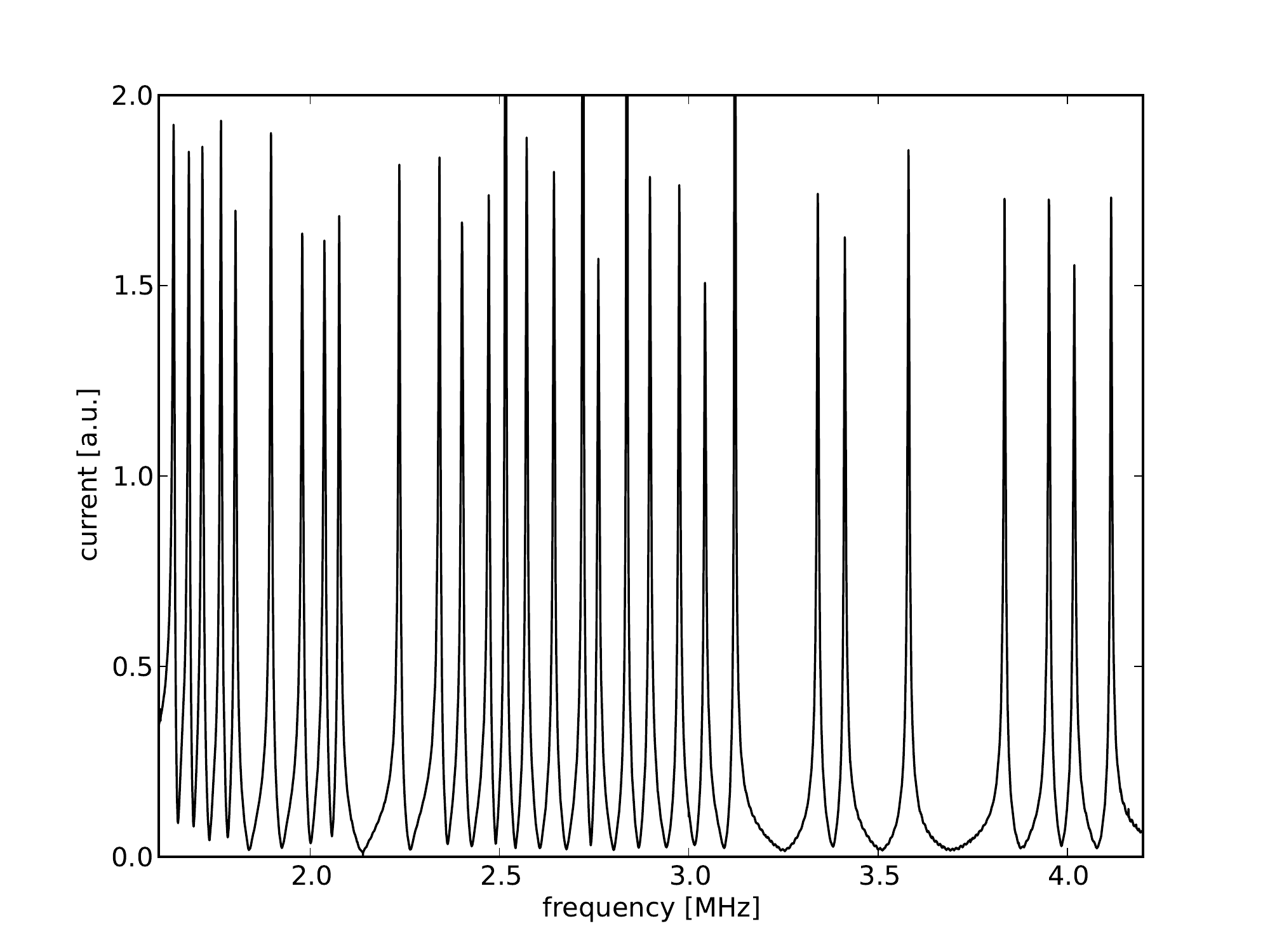}}
\subfigure[At 0.3 K where the TESes are superconducting.]{
\includegraphics[height=4.2cm]{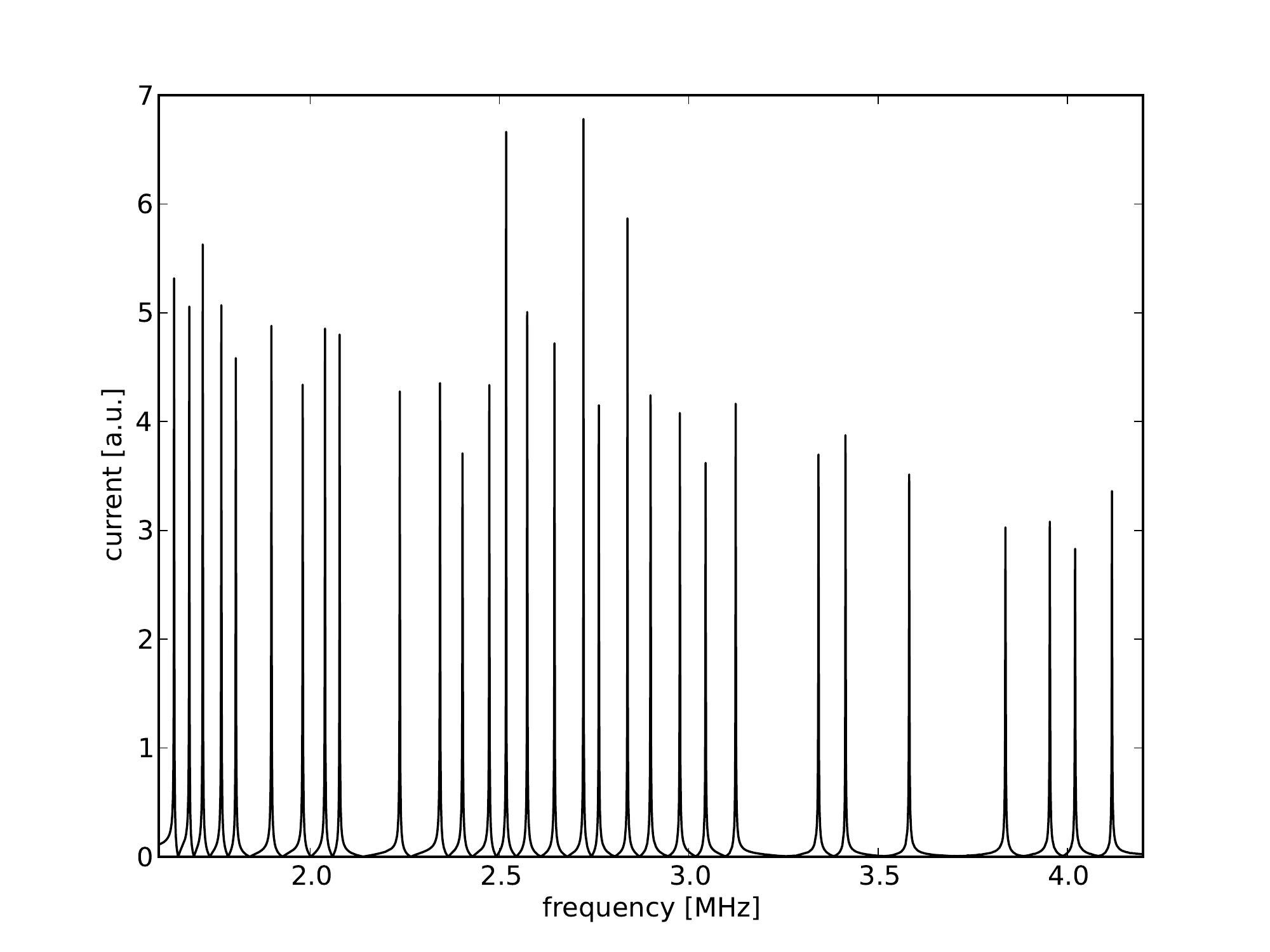}}
\end{center}
\caption{\label{fig:netanal}
The resonant peaks of LC filters
measured by frequency-sweeping a small amplitude probe.
}
\end{figure}

\begin{figure}
\begin{center}
\includegraphics[%
  width=50mm, 
  keepaspectratio]{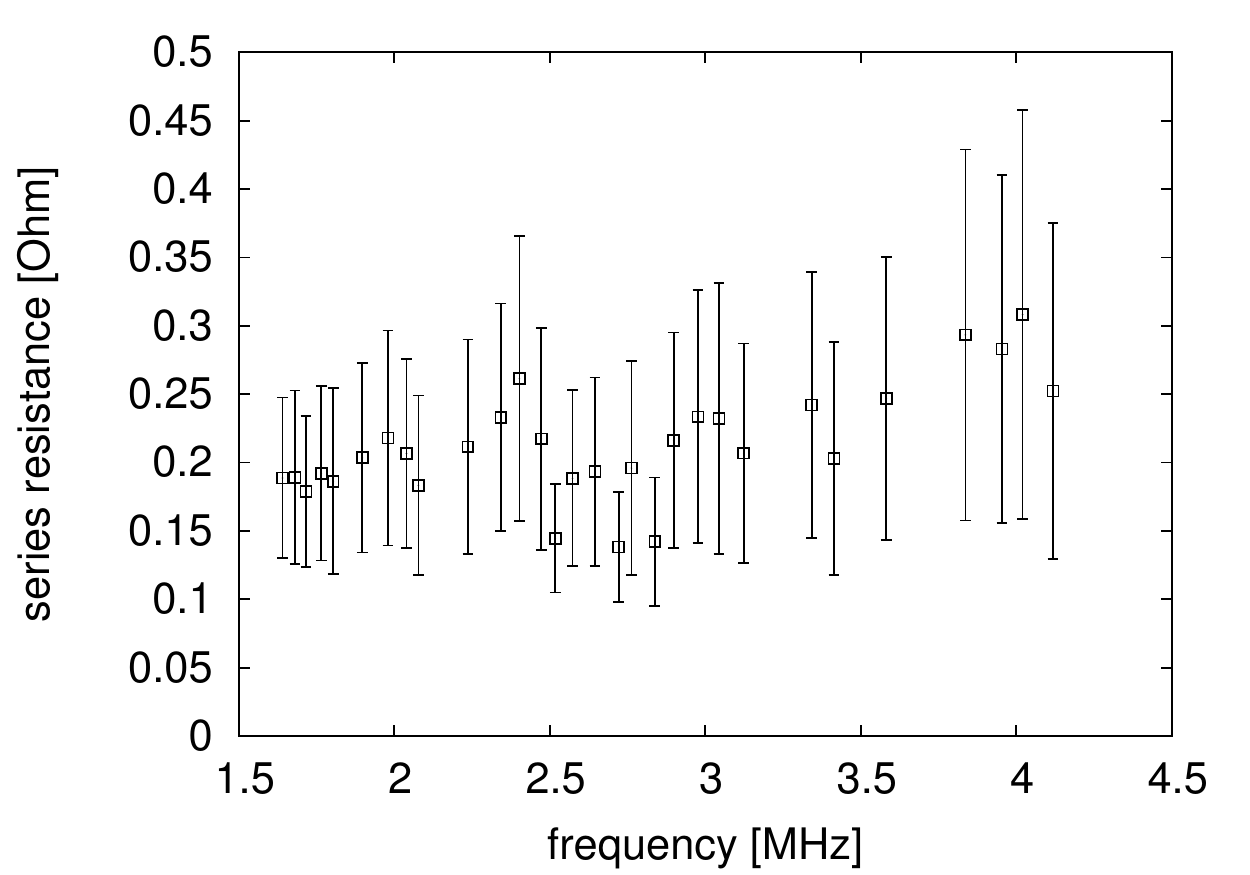}
\end{center}
\caption{Series resistance, sum of the ESR, $R_{s1}$, the wiring resistance $ R_{s0}$
and the bias resistance $R_{bias}$,  obtained by fitting the peaks in Fig. \ref{fig:netanal}(b).
}
\label{fig:Rs}
\end{figure}

\begin{figure}
\begin{center}
\subfigure[At 1.6 MHz.]{
\includegraphics[height=4cm]{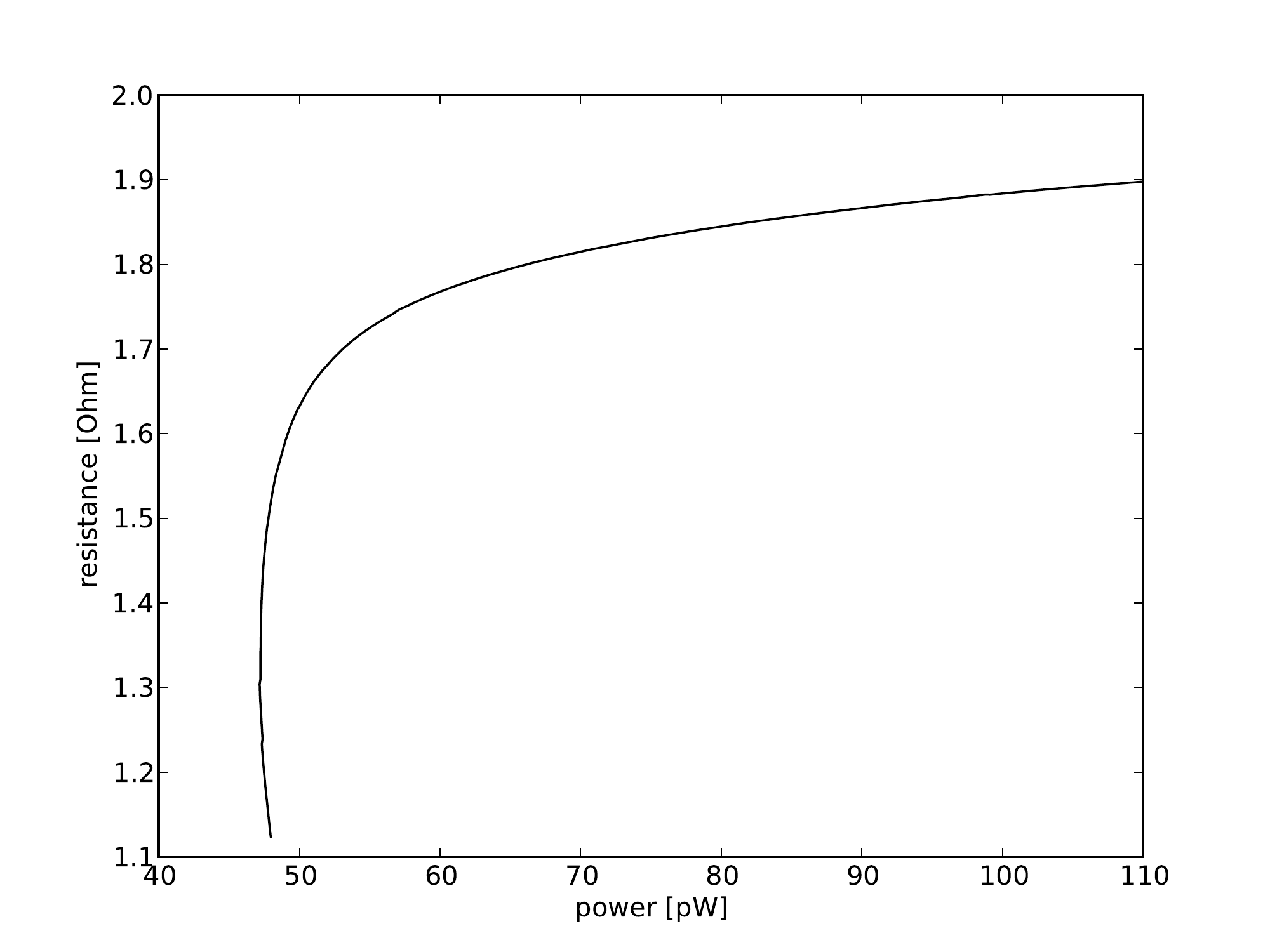}}
\subfigure[At 3.5 MHz.]{
\includegraphics[height=4cm]{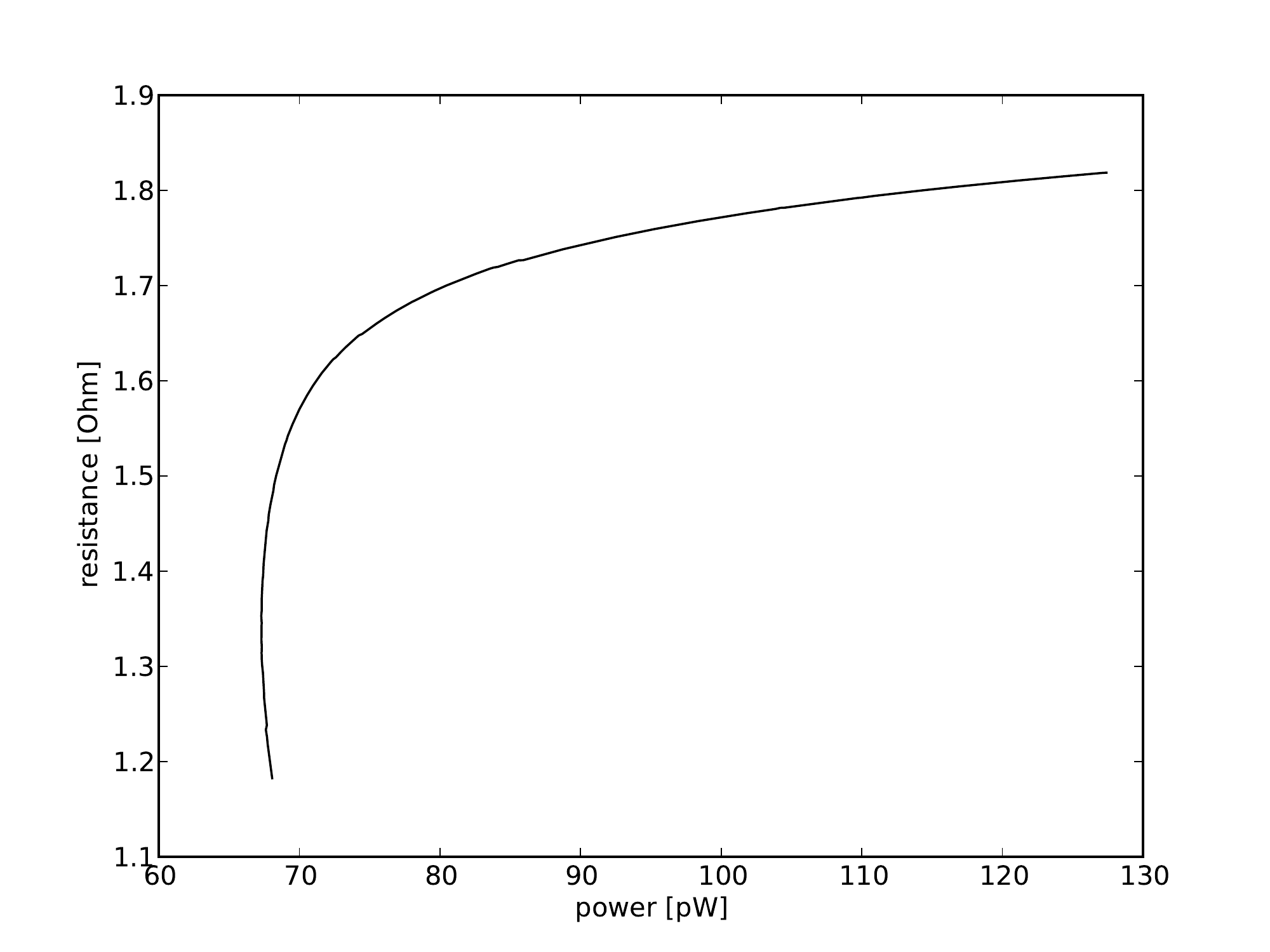}}
\end{center}
\caption{\label{fig:RP}
Bolometer resistance vs. power for voltage-biased bolometers.
}
\end{figure}

The resonant peaks of the LC filters were measured by sweeping a small amplitude probe in frequency, as shown in Fig. \ref{fig:netanal}.
We used the width of a resonant peak, $\Delta \omega = R / L$, to measure a resistance in series with an LC filter.
When the TES is superconducting, the measured resistance is equal to the sum of the series resistance,  
$R_{s0} + R_{s1} + R_{bias}$.
Figure \ref{fig:Rs} shows the series resistance obtained 
by fitting the peaks in Fig. \ref{fig:netanal}(b).
This is the sum of resistances for all the cryogenic readout electronics,
and is higher than the equivalent series resistance of the capacitors in the LC filters,
shown in \cite{ref:Kaja_LTD}.
The error bars in Fig. \ref{fig:Rs} could be originated from a fit function describing a single
LCR leg, not considering the stray capacitance, the inductance and the leakage current through 
neighboring channels in frequency spacing.
The measured resistance meets the requirements, $R_s / R_{bolo} \lesssim 0.2$,
where $R_{bolo}$ in transition is about 1 $\Omega$.
Using the new cryogenic readout, we sucessfully tuned the bolometers into the middle of transition.
Figure \ref{fig:RP} shows examples of the transition curves.

\section{Conclusion}

Controlling the series resistance and inductance is important to reduce systematic errors for CMB observations.
Even a small series resistance changes performance of the TES bolometer and creates non-linearity,
which leads to systematic errors.
We successfully reduced the series resistance using low-loss capacitors.
To reduce it further, we will implement micro-D connectors, which have a low-contact resistance
and are robust mechanically.
Though their disadvantage is a high inductance, we will use the lower-inductance stripline under development
to compensate the increase in inductance.
We have demonstrated tuning bolometers into transition at a high frequency using legacy warm electronics,
which multiplexes up to 16 channels.
New warm electronics allowing a larger multiplexing factor have been developed \cite{ref:Amy_SPIE} 
and we will test the cold readout with them. 

\begin{acknowledgements}
The \textsc{Polarbear} project is funded by MEXT KAKENHI Grant Number 21111002, 
the National Science Foundation AST-0618398, NASA grant NNG06GJ08G, and the Simons Foundation. 
The McGill authors acknowledge funding \
from the Natural Sciences and Engineering Research Council of Canada, the Canada Research Chairs Program, 
and Canadian Institute for Advanced Research. 
TES bolometers, lithographed capacitors and inductors were fabricated at the Berkeley Marvell Nanofabrication
laboratory.
\end{acknowledgements}


\end{document}